# Advancing Noise-Resilient Twist Angle Characterization in Bilayer Graphene through Raman Spectroscopy via GAN-CNN Modeling

Dan Hu[1], Ting-Fung Chung[2], Yong P. Chen[1,2,3,4,*], Yaping Qi[1,4,*]

[1]Department of Engineering Science, Faculty of Innovation Engineering, Macau University of Science and Technology, Av. Wai Long, Macau SAR, 999078, China
[2]Department of Physics and Astronomy and Elmore Family School of Electrical and Computer Engineering and Birck Nanotechnology Center and Purdue Quantum Science and Engineering Institute, Purdue University, West Lafayette, Indiana 47907, United States
[3]Institute of Physics and Astronomy and Villum Center for Hybrid Quantum Materials and Devices, Aarhus University, Aarhus-C, 8000 Denmark
[4]Advanced Institute for Materials Research (WPI-AIMR), Tohoku University, Sendai 980-8577, Japan

*Correspondence: yongchen@purdue.edu; qi.yaping.a2@tohoku.ac.jp.

**Abstract**
In this study, we introduce an innovative methodology for robust twist angle identification in bilayer graphene using Raman spectroscopy, featuring the integration of generative adversarial network and convolutional neural network (GAN-CNN). Our proposed approach showcases remarkable resistance to noise interference, particularly in ultra-low Signal-to-Noise Ratio (SNR) conditions. We demonstrate the GAN-CNN model's robust learning capability, even when SNR reaches minimal levels. The model's exceptional noise resilience negates the necessity for preprocessing steps, facilitating accurate classification, and substantially reducing computational expenses. Empirical results reveal the model's prowess, achieving heightened accuracy in twist angle identification. Specifically, our GAN-CNN model achieves a test accuracy exceeding 99.9% and a recall accuracy of 99.9%, relying on an augmented dataset containing 4209 spectra. This work not only contributes to the evolution of noise-resistant spectral analysis methodologies but also provides crucial insights into the application of advanced deep learning techniques for bilayer graphene characterization through Raman spectroscopy. The findings presented herein have broader implications for enhancing the precision and efficiency of material characterization methodologies, laying the foundation for future advancements in the field.

## 1. Introduction

Graphene has undergone extensive scrutiny as a prototypical 2D material over the past two decades, with ongoing discoveries of novel physical phenomena [1-3]. Twisted bilayer graphene (TBLG) has emerged as a focal point of investigation due to its intricate electronic properties and adjustability [4-7]. TBLG is generated by juxtaposing two single-layer graphene sheets at a specific twist angle ($\theta$), signifying a lattice rotation concerning each other. This overlay of materials with disparate lattice constants or relative twist angles results in the formation of a moiré superlattice structure [8], thereby altering the bilayer graphene's band structure and inducing distinctive characteristics.

Prior to practical applications, the precise characterization of the twist angle in TBLG is imperative. Nevertheless, currently employed conventional methodologies for twist angle identification are characterized by being time-consuming and labor-intensive. Established techniques such as Raman spectroscopy, transmission electron microscopy, and low-energy electron diffraction necessitate specialized manual input for data interpretation. Conversely, machine learning techniques, encompassing deep learning methods, have exhibited efficacy in facilitating Raman spectroscopy for the expeditious identification of twist angles in TBLG [9-13].

Despite the promising outcomes attained by current methods, there persists a reliance on conventional analyses as an initial step in these approaches. This involves acquiring the primary Raman characteristic peaks of the sample, specifically the G and 2D peaks for graphene. Subsequent analysis is conducted based on parameters such as intensity, Raman shift, and full width at half maximum for these identified peaks. The values resulting from this meticulous analysis are subsequently fed into a learning model for predictive purposes. This underscores the fact that conventional machine learning methods typically necessitate the development of finely tuned and reliable features. Such features demand the involvement of personnel possessing relevant experience and expertise to ensure the efficacy of the overall approach.

As of now, there has been a dearth of investigations establishing a direct, unequivocal correlation between Raman spectroscopy and the twist angles inherent in bilayer graphene. Deep learning techniques, particularly exemplified by Convolutional Neural Networks (CNNs), which inherently feature automatic feature extraction and an end-to-end model architecture, present a highly pertinent approach for tackling this research challenge. This paper introduces a novel methodology that integrates Raman spectroscopy of bilayer graphene with a deep learning model founded on Generative Adversarial Networks (GANs) and CNNs (GAN-CNN) to discern and identify twist angles with enhanced precision.

This investigation delves into the potential of integrating Raman spectroscopy with deep learning methodologies to discern the twist angle of bilayer graphene. The objective is to leverage the capabilities of deep learning algorithms with the aim of improving the accuracy and efficiency of twist angle identification. A comprehensive suite of algorithms is considered for comparison, encompassing traditional machine learning methods such as Logistic Regression (LS), Decision Tree (DT), Random Forest (RF), Support Vector Machine (SVM), and K-Nearest Neighbor (KNN). Additionally, deep learning techniques, including Artificial Neural Networks (ANN), Convolutional Neural Networks (CNN), and Generative Adversarial Networks (GAN), are evaluated for their efficacy in this context. These algorithms present diverse approaches to analyze Raman spectroscopy data, facilitating the extraction of meaningful insights for twist angle identification.

The findings of this study are anticipated to enrich the overarching comprehension of integrating deep learning within the realm of Raman spectroscopy for materials analysis, particularly with a targeted emphasis on discerning the twist angle of bilayer graphene. The implications of the results extend beyond the immediate scope, potentially serving as a foundational milestone for subsequent advancements. Moreover, these outcomes are poised to open avenues for broader applications in materials science and related domains, thereby contributing to the continual evolution of analytical methodologies and technological applications in these fields.

**2. Dataset and preprocessing**

The graphene Raman spectra utilized in this study were derived from authentic experimental data conducted by Dr. Ting-Fung Chung at the Department of Physics and Astronomy, Purdue University. Raman imaging was executed on graphene samples affixed to $SiO_2$/Si substrates, employing a 532 nm excitation wavelength. The dataset encompassed both single-layer graphene (SLG) and bilayer graphene samples, featuring twist angles within the ranges of 0-9°, 9-20°, and 20-30°, thereby constituting four distinct categories. The dataset consisted of a total of 209 Raman spectra, as delineated in Figure 1 (a), with the respective distribution of data. Each Raman spectrum comprised 1312 intensity values within the wavenumber range of 1200 $cm^{-1}$ to 3000 $cm^{-1}$. Figure 1(b) illustrates the distinct categories of Raman spectra, highlighting the subtle differences existing between various twist angles within the dataset. This comprehensive dataset serves as the foundation for the subsequent analysis and application of deep learning methodologies in the context of twist angle identification.

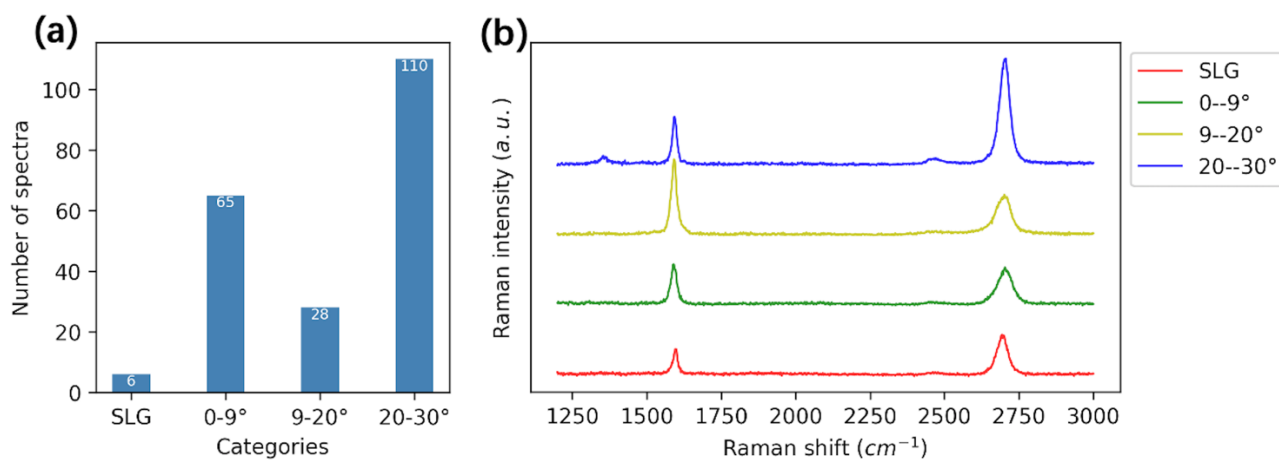

Figure 1 (a). The graphical representation showcases the distribution of graphene Raman spectra across different categories. (b). The figures depict Raman spectra of graphene characterized by various twist angles.

The collected Raman spectral data exhibited considerable variation in intensity. If employed directly for training the identification model, features with larger values could potentially overshadow those with smaller values, undermining the robustness of the training model. To mitigate the influence of significant disparities in the data, preprocessing steps were deemed necessary to rectify the comparability issue inherent in Raman spectral data. In this investigation, a normalization process was applied to ensure that all data fell within the [0,1] range, as articulated in Equation 1:

$$x^* = \frac{x - min}{max - min} \tag{1}$$

Here, in Equation 1, x signifies the input, while x∗ represents the normalized data. The terms 'min' and 'max' denote the minimum and maximum values of the input data, respectively. Normalizing the spectral data within a defined range was instrumental in enhancing the efficiency of the model training process. The Raman spectral curves after normalization are visually depicted in Figure 2.

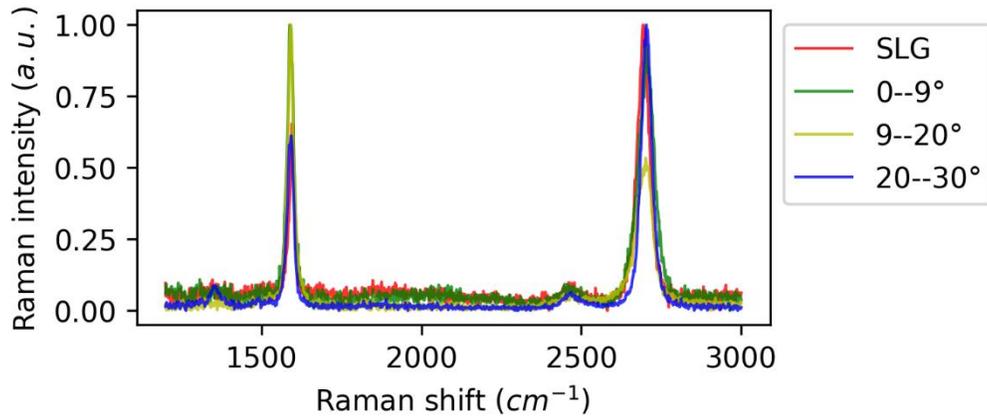

Figure 2 Diagrammatic representation illustrating the normalized Raman spectrum.

## 3. GAN-CNN based twist angle identification model

In practical applications, the challenge arises when attempting to discern weak signals of trace substances from the substrate background using Raman analysis techniques. This complexity hinders the observation of the target substance's signal and imposes limitations on the quantity of spectral data attainable through actual experiments [14]. Notably, the efficacy of classification models is often contingent on the dataset size. While deep learning networks have demonstrated considerable success across various domains, they necessitate substantial datasets to optimize network parameters and mitigate overfitting concerns [15]. To address the issue of data scarcity, one approach involves the repetition of experiments to generate new data, an endeavor demanding considerable human and material resources. Alternatively, data augmentation presents a viable method to overcome overfitting challenges and enhance the accuracy of classification algorithms [16]. To tackle the challenge, this study adopts a deep convolutional GAN data augmentation technique. This method is employed to generate a multitude of independent and identically distributed samples from the original Raman spectral dataset. Subsequently, it is integrated with a one-dimensional Convolutional Neural Network (1D CNN) algorithm to formulate a model designed for the identification of twist angles in bilayer graphene. Referred to as the GAN-CNN model in this study and depicted in Figure 3, its purpose is to establish a scientific foundation for the utilization of Raman spectroscopy technology in conjunction with deep learning methods for graphene analysis.

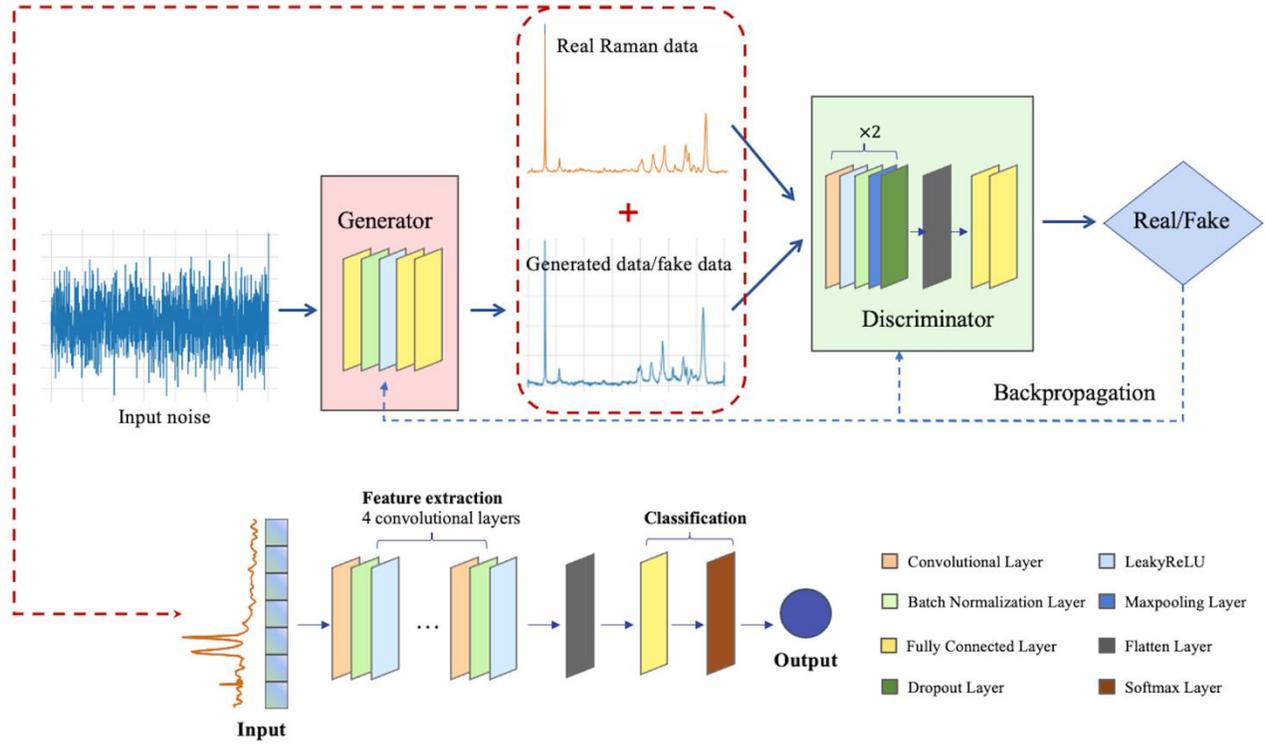

Figure 3 Comprehensive depiction of the GAN-CNN model framework.

## 3.1 Generative adversarial neural networks

The GAN architecture, initially proposed by Goodfellow et al. [17], operates as a framework for generative modeling through adversarial training, as depicted in the conceptual diagram in the upper section of Figure 3. GAN comprises two essential components: the generator and the discriminator. The discriminator functions as a binary classifier tasked with determining whether an input is derived from real data or generated data. Specifically, the discriminator assigns the label y = 1 to samples originating from real data and y = 0 to those from the generator. The primary objective of the discriminator is to furnish the probability that an input sample x is drawn from the true distribution, denoted as p (y = 1|x). Its objective function involves minimizing cross-entropy, as defined in Equation 2:

$$\min(E_{x \sim p_{\text{data}}}[y \log p(y = 1|x) + (1 - y) \log p(y = 0|x)]) \qquad (2)$$

Simultaneously, the generator is tasked with capturing the distribution of sample data, with the goal of having the discriminator classify the generated samples as real. This objective is encapsulated in Equation 3, where z denotes the input noise variable, and E represents the expected operation.

$$\min(E_{z \sim p_z}[logD(G(z))]) \tag{3}$$

The training process of the GAN can be conceptualized as a concurrent optimization of the generator and discriminator parameters. This aims to minimize and maximize, respectively, the objective expressed in Equation 4:

$$\min_G \max_D V(D, G) = E_{x \sim p_{\text{data}}(x)}[logD(x)] + E_{z \sim p_z(z)}\left[log\left(1 - D(G(z))\right)\right] \tag{4}$$

The objectives of the two networks are inherently opposing, perpetually engaging in adversarial training to enhance performance [18]. Drawing from game theory principles, when the discriminator believes that the probability of all input samples being real is 50%, the two network models can theoretically reach Nash equilibrium. At this juncture, the generator can be regarded as having successfully learned the authentic distribution of real data.

### 3.2 Construction of Neural Networks and Training Setup

The architecture of the GAN-CNN model involves three distinct neural networks: the generator, discriminator, and classification networks. In this study, a straightforward fully connected neural network was established as the generator model. This generator comprises two fully connected layers utilizing LeakyReLU as the activation function. Gaussian noise with dimensions matching the Raman spectra data serves as the input for the generator network. The generated data is produced through fully connected layers, incorporating LeakyReLU activation function and Batch Normalization. This process results in Raman spectra data possessing similar features to the original input data. Concurrently, the discriminator network is constituted by a two-layer 1D CNN. This network takes Raman spectra as input and employs a max-pooling layer after each convolutional layer to down sample the input to the dimensions required by the original Raman signal. Ultimately, the discriminator network outputs a binary value through two fully connected layers. The detailed structures of the generator and discriminator networks are illustrated in Figure 4.

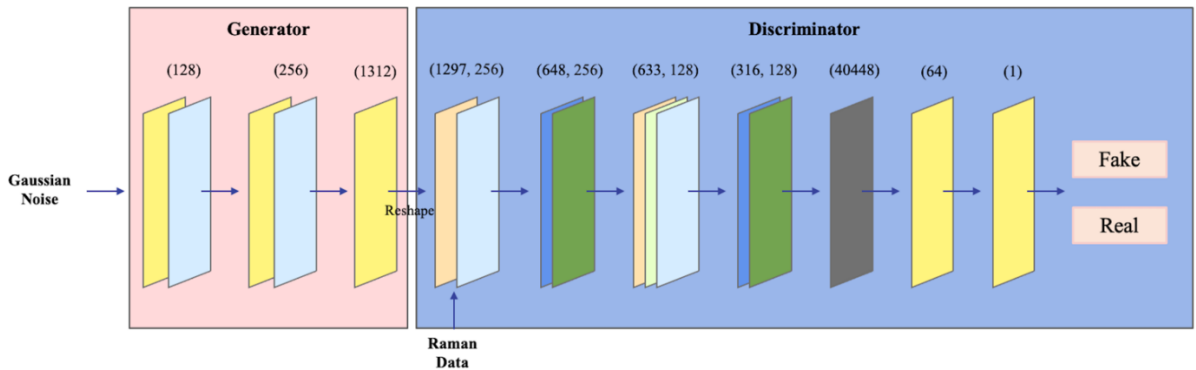

Figure 4 Architectures of the neural networks, illustrating both the generator and discriminator components.

Given that the training of GAN can be perceived as an ongoing evaluation of the authenticity of generated data, the binary cross-entropy loss serves as the selected loss function for calculating the loss during training, as represented in Equation 5:

$$loss = -\frac{1}{m}\sum_{i=1}^{m}[y_i log\hat{y}_i + (1-y_i)log(1-\hat{y}_i)] \qquad (5)$$

In Equation 5, m represents the number of samples, y denotes the correct label value (where the positive class value is 1, and the negative class value is 0), and $\hat{y}$ signifies the predicted probability value (with y belonging to the range (0, 1)). The loss function quantifies the disparity between the actual sample label and the predicted probability. The generator utilizes the Adam optimizer, while the discriminator employs the RM-Sprop optimizer, as recommended by Arjovsky et al. [19], given its efficacy in handling non-stationary problems. The learning rate is set at 0.0001. The optimizer iteratively adjusts the neural network weights based on the training data, ensuring the model's adaptability. The dataset is split into a 4:1 ratio for the training set and test set, respectively. Raman spectral data augmentation is conducted using GAN for graphene samples featuring four distinct twist angles. Following data augmentation, each category is expanded with an additional 1000 spectra, as detailed in Table 1.

Table 1 Information about the datasets.

| Original Dataset | | Data Augmentation | | Augmented Dataset | |
|---|---|---|---|---|---|
| Category | Number | Category | Number | Category | Number |
| SLG | 6 | SLG | 1000 | SLG | 1006 |
| 0-9° | 65 | 0-9° | 1000 | 0-9° | 1065 |
| 9-20° | 28 | 9-20° | 1000 | 9-20° | 1028 |
| 20-30° | 110 | 20-30° | 1000 | 20-30° | 1110 |
| **Total**: | 209 | **Total**: | 1000 | **Total**: | 4209 |

The Raman data, both before and after augmentation, is visually presented in Figure 5. It is noteworthy that the generated Raman data may exhibit some degree of noise and transformation, yet this does not affect the feature peak values. Such noise and transformations can be viewed as perturbations to the original data, contributing to the training of a more robust model capable of handling practical applications involving noise and transformations.

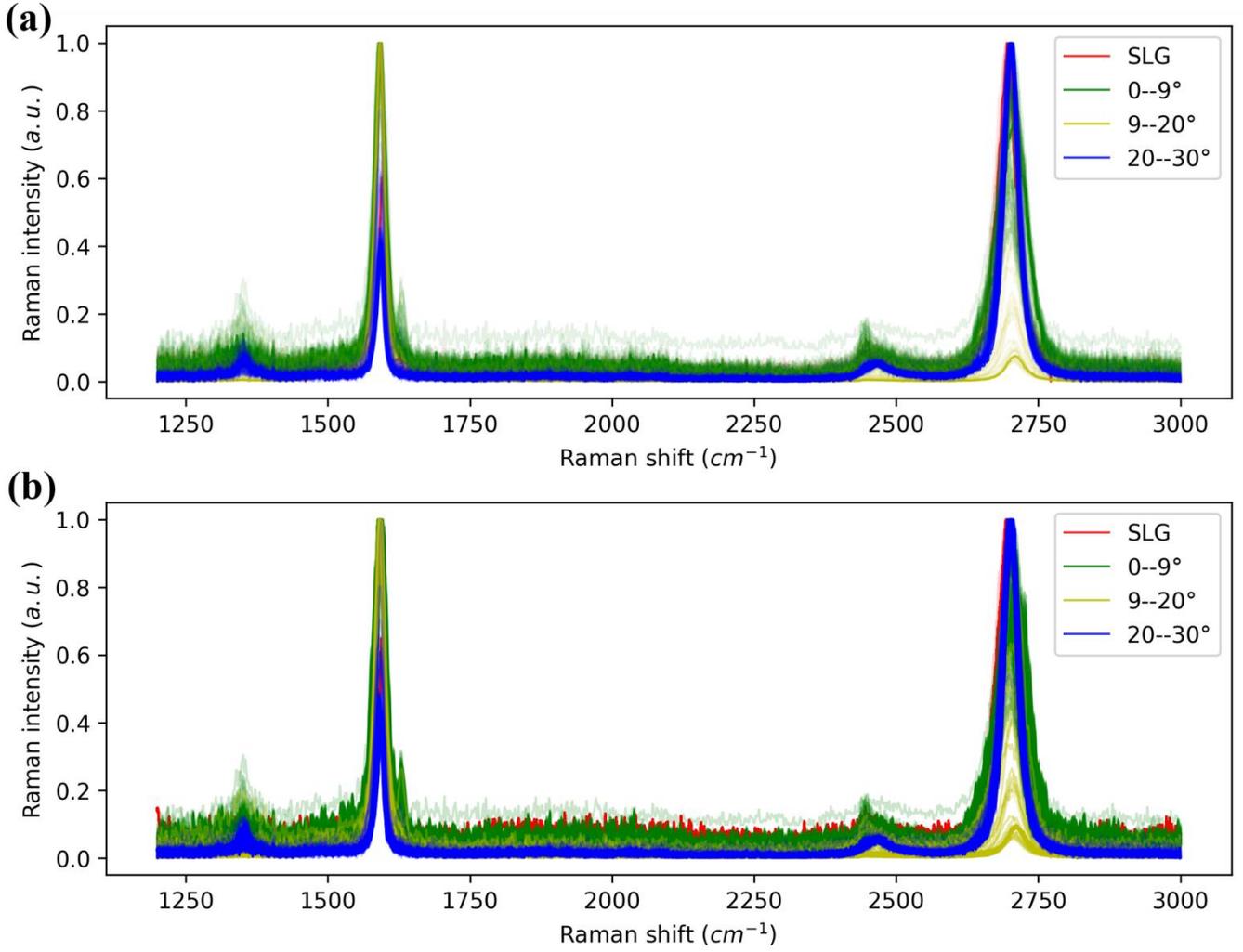

Figure 5 (a) Graphical representation of raw graphene spectral data. (b) Graphical depiction of augmented graphene spectral data.

The neural network architecture, along with the specific parameters utilized for classification, is illustrated in Figure 6. The model comprises four convolutional layers designed to extract data features, succeeded by a flattened layer that channels data into a fully connected layer. The SoftMax activation function is employed for classification purposes. The classifier utilizes sparse categorical cross-entropy to compute the loss, as defined in Equation 6, where m represents the number of samples, and k denotes the number of sample categories.

$$loss = -\frac{1}{m}\sum_{i=1}^{m}\sum_{j=1}^{k} y_{ij} \log \hat{y}_{ij} \qquad (6)$$

The classifier outputs the same number as the categories in the dataset. Additionally, the classifier employs the Adam optimizer with a learning rate set at 0.0001, chosen for its simplicity and effectiveness.

Throughout the model training process, the GAN-CNN utilizes the augmented dataset, whereas the CNN exclusively employs the original dataset (Table 1). The training and test sets for all models are partitioned in a 4:1 ratio.

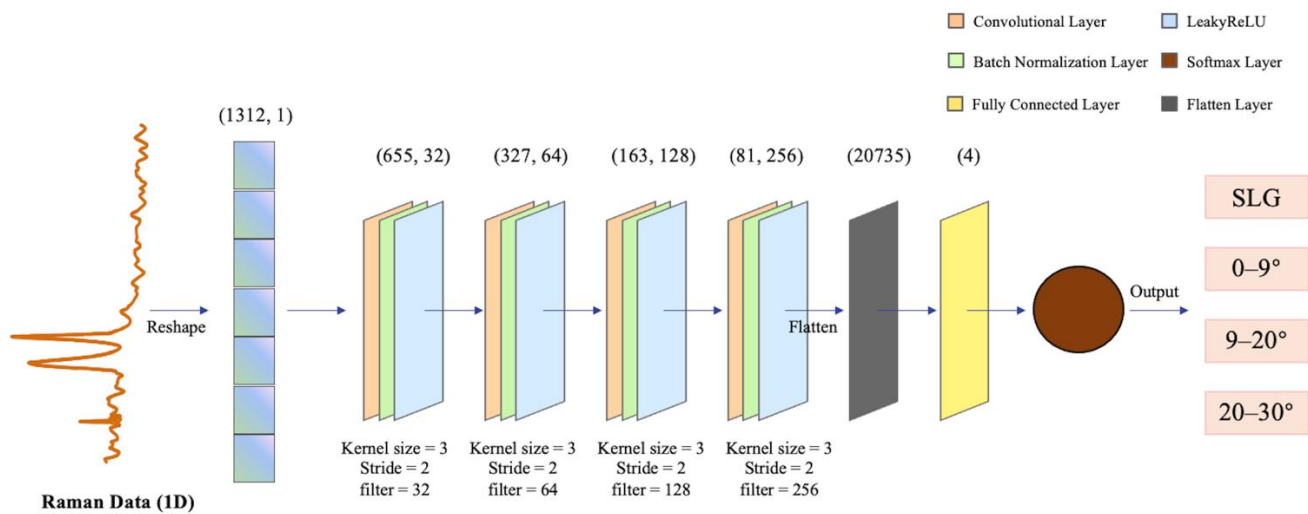

figure 6 Illustration of the architecture of the four-layer CNN employed for Raman spectroscopy classification.

In the context of a deep learning network model, the intermediate layers encapsulate abundant spatial and semantic information. To illustrate the feature extraction process of the classifier, this experiment employs gradient-weighted class activation mapping (Grad-CAM) to pinpoint and illustrate the decisions by deep neural networks. The visualization in Figure 7 elucidates the features learned by the network in the convolutional layer.

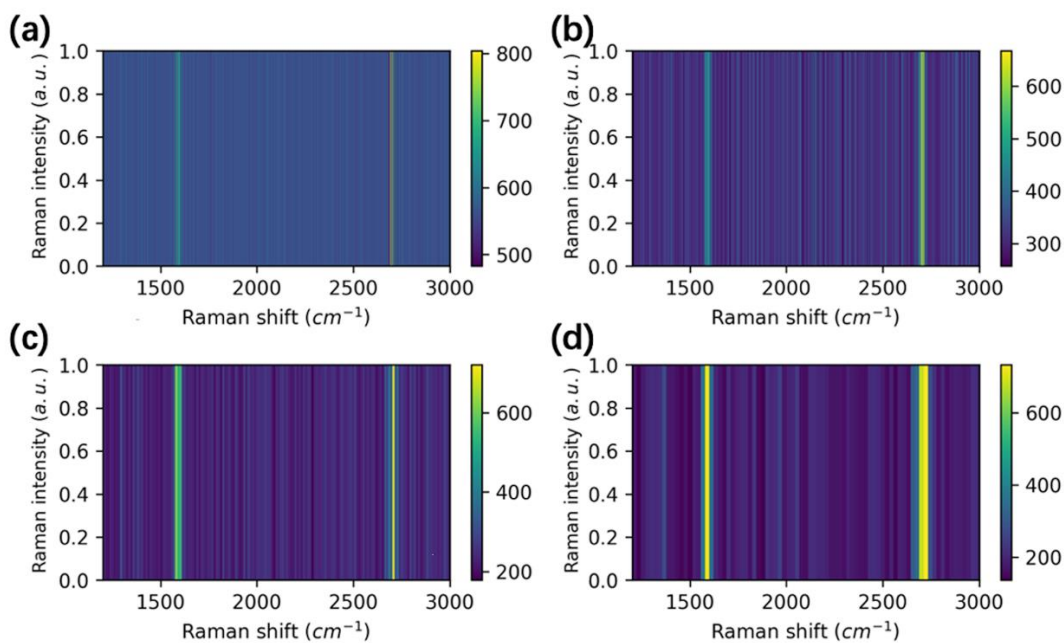

Figure 7 Mapping of activations for the (a) first, (b) second, (c) third, and (d) fourth convolutional layers.

The highlighted heatmap areas in the image serve as the basis for the model's output determination. In essence, if a Raman spectrum exhibits a robust feature peak in a particular region, and this significant peak is absent in another spectrum of the same class, the activation value of the feature map associated with this strong peak will carry higher weight during the forward propagation process. Consequently, the corresponding region in the heat map will appear brighter. As depicted in Figure 7, the model, with the progression of convolutional layers, adeptly extracts the positions of robust Raman peaks in the Raman spectra, showcasing its ability to discern salient features accurately.

4. Experiments and Analysis

To assess the efficacy of the GAN-CNN model enhancement module, both the original data and newly generated sample data using GAN were visualized using the t-distributed stochastic neighbor embedding (t-SNE) method [20]. T-SNE, a nonlinear dimensionality reduction technique, facilitates the mapping of high-dimensional data onto a scatter plot of low-dimensional data while preserving the structure of the data information.

As illustrated in Figure 8(a), the distribution of the original data appears scattered due to the small sample size, making it challenging to distinguish. In contrast, Figure 8(b) reveals that the data-enhanced samples are more discernible compared to the original samples. The data augmentation not only expands the coverage of the dataset but also preserves the distribution of key features present in the original data.

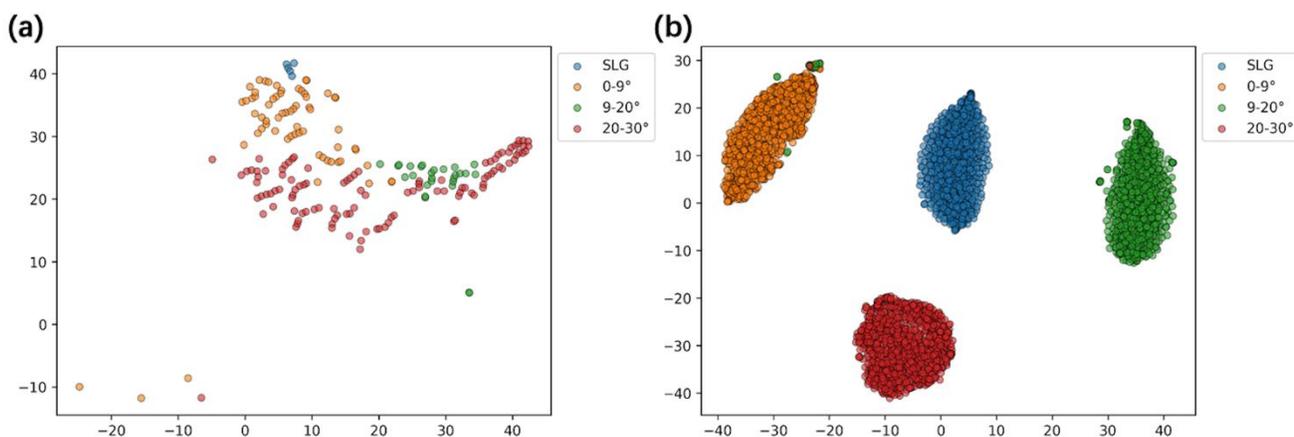

Figure 8 Presentation of (a) the original and (b) augmented datasets featuring graphene with various twist angles using t-SNE visualization.

In the training process of deep learning models, the number of iterations directly affects the in the training process of deep learning models, the number of iterations significantly influences the model's performance. Figure 9 illustrates the relationship between classification accuracy and loss values for different iteration numbers during the model training process, both with and without GAN enhancement.

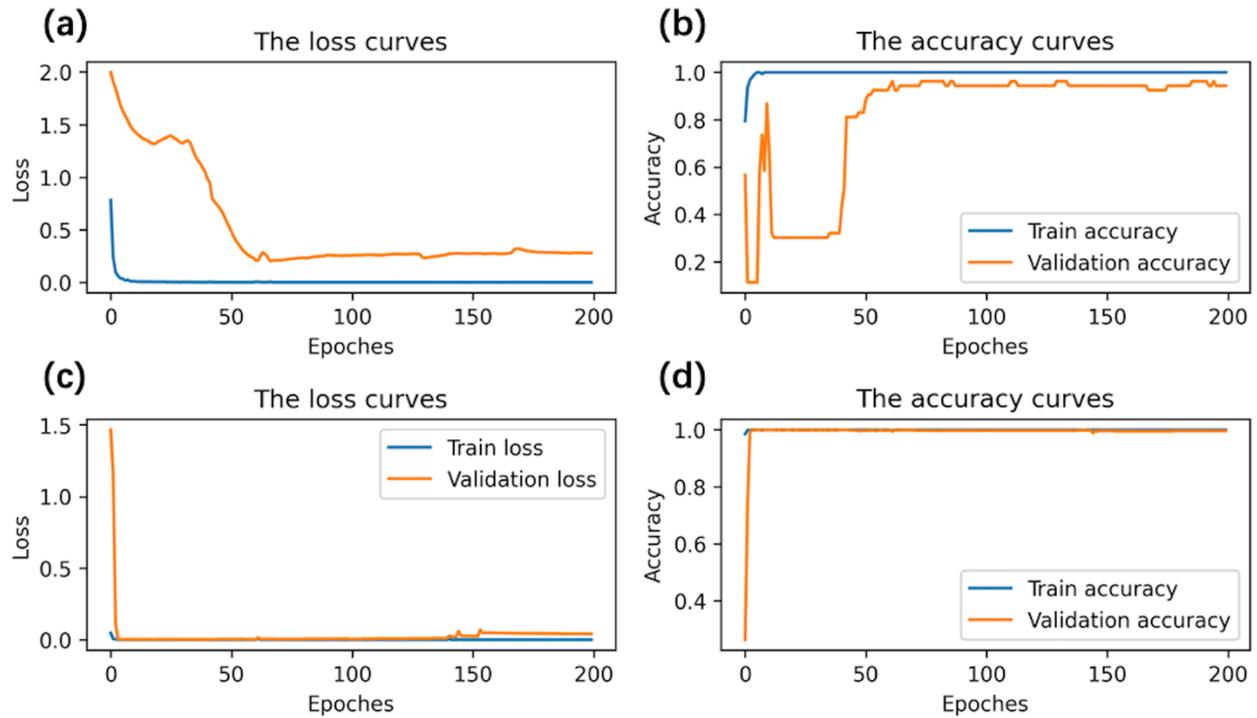

Figure 9 (a-b) Graphs illustrating loss curves and accuracy curves for training and validation data during the CNN model training process. (c-d) Graphs depicting loss curves and accuracy curves for training and validation data during the GAN-CNN model training process.

The model underwent training on the training set for 200 epochs, utilizing a batch size of 16. Notably, the loss function values for both the training set and the validation set exhibit a similar descending trend, steadily decreasing over time. However, GAN-CNN achieves effective convergence at a faster rate. Various metrics, including accuracy, precision, recall, and mean accuracy, were assessed through ten-fold cross-validation to gauge the performance of the multi-class network model. Additionally, the model's identification performance in each category was evaluated using the confusion matrix, as depicted in Figure 10.

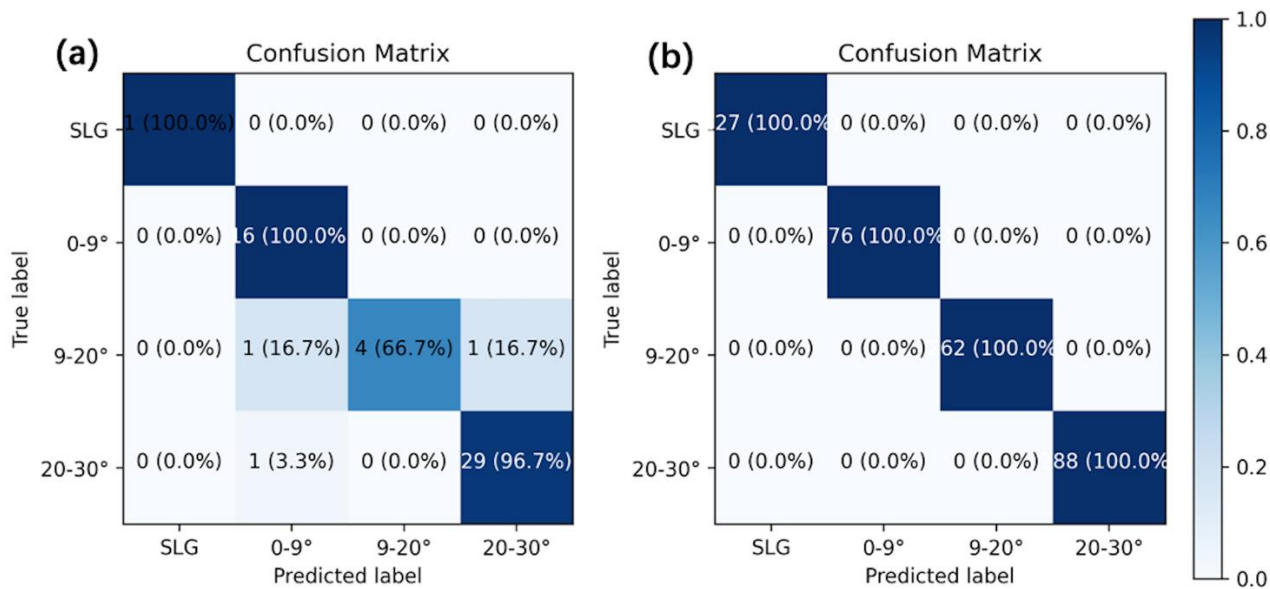

Figure 10 presents the confusion matrices for the CNN classification model under two conditions: (a) without the inclusion of GAN, and (b) with the incorporation of GAN.

The training results obtained using the original dataset and the GAN-generated dataset are presented in Table 2. The CNN model demonstrates commendable accuracy at 94.3%, affirming the applicability of the neural network model designed in this study for aiding in the analysis of Raman spectroscopy. Notably, the GAN-CNN model, incorporating data augmentation, further enhances accuracy to 100%, underscoring the substantial performance improvement afforded by the augmentation module in the classifier model.

Table 2 The classification performance of CNN and GAN-CNN models.

| Method | Accuracy | Recall | Precision | 10-fold Cross-validation |
|---|---|---|---|---|
| CNN | 0.943 | 0.908 | 0.964 | 0.962 |
| GAN-CNN | 1.000 | 1.000 | 1.000 | 0.998 |

To substantiate the advantages of the GAN-CNN model over traditional analysis methods in facilitating Raman spectroscopy analysis of TBLG, experiments were conducted to compare its classification performance with five classic conventional machine learning algorithms. These include RF, SVM, DT, KNN, LR, and an ANN model with two hidden layers. Table 3 presents the accuracy, precision, recall on the test set, as well as the mean accuracy of ten-fold cross-validation for GAN-CNN and all the compared methods. For a more visual comparison of the experimental results, refer to Figure 11.

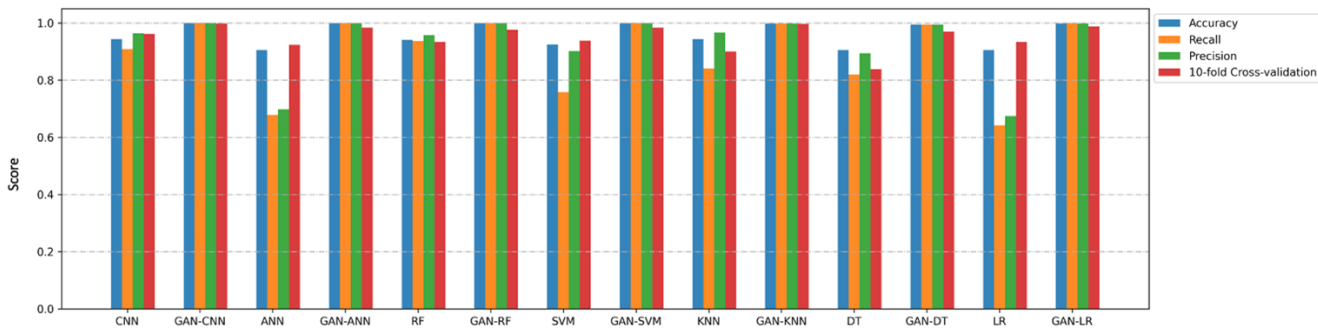

Figure 11 Bar chart presenting a comparative analysis of the performance of various models in the classification of bilayer graphene at different twist angles.

The experimental results showcase that the GAN-CNN model outperformed other methods, achieving 100% accuracy on the test set and 99.8% accuracy through ten-fold cross-validation. Notably, the classification performance of the five machine learning models and the ANN model experienced a significant boost after incorporating GAN enhancement. In comparison to methods that do not leverage GAN, the four-layer CNN model employed in this research emerges as superior to other algorithms in accurately identifying the twist angles of TBLG.

Table 3 Evaluation of the classification performance of diverse models.

| Method | Accuracy | Recall | Precision | 10-fold Cross-validation |
|---|---|---|---|---|
| CNN | 0.943 | 0.908 | 0.964 | 0.962 |
| GAN-CNN | 1.00 | 1.00 | 1.00 | 0.998 |
| ANN | 0.906 | 0.678 | 0.698 | 0.924 |
| GAN-ANN | 0.999 | 0.999 | 0.999 | 0.983 |
| RF | 0.941 | 0.938 | 0.958 | 0.933 |
| GAN-RF | 0.999 | 0.999 | 0.999 | 0.976 |
| SVM | 0.925 | 0.758 | 0.902 | 0.938 |
| GAN-SVM | 0.999 | 0.999 | 0.999 | 0.984 |
| KNN | 0.943 | 0.841 | 0.966 | 0.900 |
| GAN-KNN | 0.998 | 0.998 | 0.998 | 0.996 |
| DT | 0.906 | 0.819 | 0.893 | 0.838 |
| GAN-DT | 0.994 | 0.994 | 0.994 | 0.970 |
| LR | 0.906 | 0.642 | 0.674 | 0.933 |
| GAN-LR | 0.998 | 0.999 | 0.998 | 0.988 |

In the process of collecting Raman spectroscopy, various uncontrollable interferences, such as noise from equipment instruments and the environment, fluorescence, etc., may be present. The existence of multiple interferences imposes higher demands on the stability of the spectral classification model. Consequently, the model should exhibit enhanced anti-noise interference capabilities and robustness to more complex detection scenarios. The dataset of TBLG acquired in this study has undergone artificial processing and classification (Chung et al. [21]). To validate the robustness of the constructed Raman spectroscopy classification model and its potential practical applications for optimal rapid identification of the twist angles of BLG, the experiment introduced Gaussian noise with different signal-to-noise ratios (SNRs) to the spectral data, forming the test set. This simulation replicates the influencing factors in actual experiments and analyzes the impact of noise on the model. The SNR calculation formula is expressed as follows in Equation 7:

$$SNR = 10 log_{10} \frac{P_s}{P_n} = 10 log_{10} \frac{\sum x^2(t)}{\sum n^2(t)} \quad (7)$$

The signal-to-noise ratio (SNR) is calculated using the formula presented in Equation 7, where x(t) and n(t) denote the signal and noise, respectively. SNR represents the ratio of normal signal power to noise signal power in the absence of a signal, measured in decibels (dB). A lower SNR indicates greater noise, and an SNR of 0 dB signifies equal signal and noise power. In this comparative experiment, Gaussian noise with SNRs of 0 dB and 30 dB was introduced to Raman spectra as the test set, simulating raw data that has not undergone processing in actual experiments. The comparison of Raman spectra before and after adding noise is depicted in Figure 12.

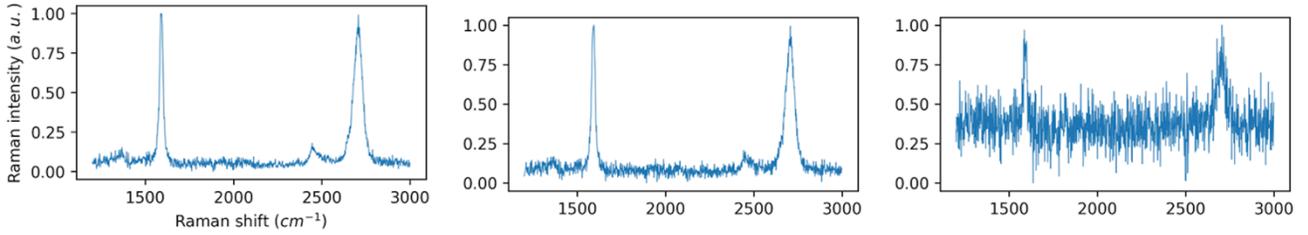

Figure 12 Raman spectra with Gaussian noise. Sequentially from left to right: the original Raman spectrum, the Raman spectrum with SNR of 30 dB, and the Raman spectrum with SNR of 0 dB.

Analyzing the effects of different noise intensities on the identification accuracy of various machine learning models, the experimental results are detailed in Table 4. In Table 4, the GAN-CNN model showcases robust classification performance for Raman spectra at signal-to-noise ratios (SNR) of both 30 dB and 0 dB, achieving accuracies of 100% and 99.6%, respectively.

Table 4. Comparative analysis of model classification performance under varying SNR.

| Method | SNR=0dB | | | SNR=30dB | | |
| --- | --- | --- | --- | --- | --- | --- |
| | Accuracy | Recall | Precision | Accuracy | Recall | Precision |
| CNN | 0.906 | 0.690 | 0.764 | 0.943 | 0.741 | 0.711 |
| GAN-CNN | 0.996 | 0.996 | 0.996 | 1.000 | 1.000 | 1.000 |
| ANN | 0.302 | 0.250 | 0.075 | 0.660 | 0.527 | 0.638 |
| GAN-ANN | 0.443 | 0.451 | 0.255 | 0.732 | 0.746 | 0.614 |
| RF | 0.302 | 0.250 | 0.078 | 0.792 | 0.768 | 0.767 |
| GAN-RF | 0.376 | 0.379 | 0.263 | 0.857 | 0.863 | 0.896 |
| SVM | 0.302 | 0.359 | 0.113 | 0.340 | 0.411 | 0.455 |
| GAN-SVM | 0.311 | 0.321 | 0.151 | 0.817 | 0.826 | 0.865 |
| KNN | 0.302 | 0.250 | 0.075 | 0.321 | 0.478 | 0.420 |
| GAN-KNN | 0.253 | 0.250 | 0.063 | 0.721 | 0.734 | 0.848 |
| DT | 0.717 | 0.440 | 0.374 | 0.792 | 0.592 | 0.637 |
| GAN-DT | 0.579 | 0.577 | 0.508 | 0.845 | 0.842 | 0.888 |
| LR | 0.302 | 0.250 | 0.075 | 0.509 | 0.455 | 0.598 |
| GAN-LR | 0.301 | 0.311 | 0.244 | 0.995 | 0.995 | 0.995 |

For a more intuitive assessment of the anti-noise interference ability of different classification algorithms, refer to Figure 13.

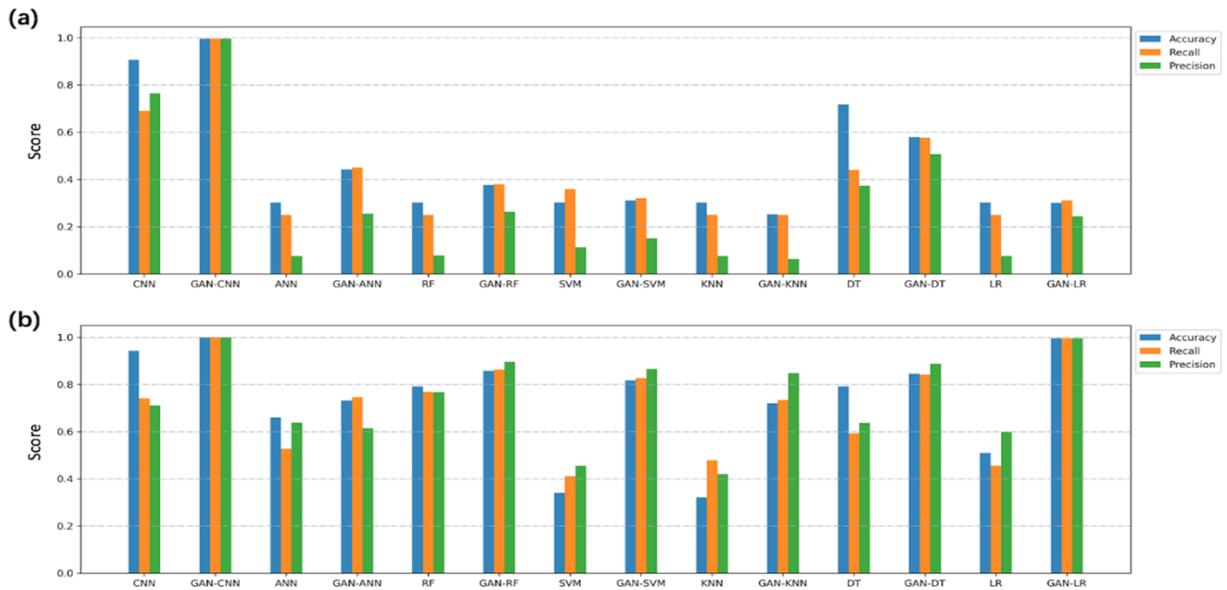

Figure 13 Bar charts illustrating the classification performance of diverse models under distinct SNR: (a) SNR=0 dB, (b) SNR=30 dB.

A comparative analysis of the GAN-CNN model against other models lacking GAN enhancement (Figure 11) underscores the GAN-CNN's superior resilience to noise interference. Conversely, models without the

GAN enhancement module exhibit significantly diminished performance, with the CNN model experiencing around a 20% decline in both recall and precision. For spectra with an SNR of 30 dB, SVM, KNN, and LR accuracies are all below 60%. Additionally, these models demonstrate a diminishing performance trend as SNR decreases. At an SNR of 0 dB, the average accuracy of all comparison models, excluding CNN, falls below 50%, indicating their ineffectiveness in performing identification tasks under such challenging noise conditions.

All GAN-augmented models exhibit an average accuracy exceeding 70% when the SNR is relatively high. Notably, LR experiences the most substantial improvement, achieving an accuracy of 99.5%. However, in scenarios where the signal and noise are at comparable levels, making features challenging to distinguish, all compared models are deemed to lack classification ability. This observation underscores the importance of preprocessing and feature extraction when employing conventional machine learning methods for spectral data analysis. While traditional machine learning, such as the LR model, can yield satisfactory results for identifying spectra with apparent features and sufficient sample size, they often display instability and high sensitivity to environmental factors when tasked with classifying raw Raman spectra without preprocessing. This limitation becomes evident, particularly when SNR is 0, emphasizing the reliance of traditional machine learning methods on preprocessing and their inferior ability to analyze raw Raman spectra compared to CNNs based on deep learning.

The experimental outcomes affirm the exceptional noise resistance of the proposed GAN-CNN model in identifying the twist angles of bilayer graphene through Raman spectroscopy. Figure 14 presents the Grad-CAM results extracted from the fourth convolutional layer of the classification model under an SNR of 0. This visualization demonstrates that even in ultra-low SNR conditions, the model adeptly learns spectral features, underscoring its robust learning ability in complex environments. Moreover, these findings validate that the GAN-CNN model can directly utilize collected Raman spectra as input without necessitating preprocessing steps. This streamlined approach not only ensures accurate classification but also contributes to time and computational efficiency.

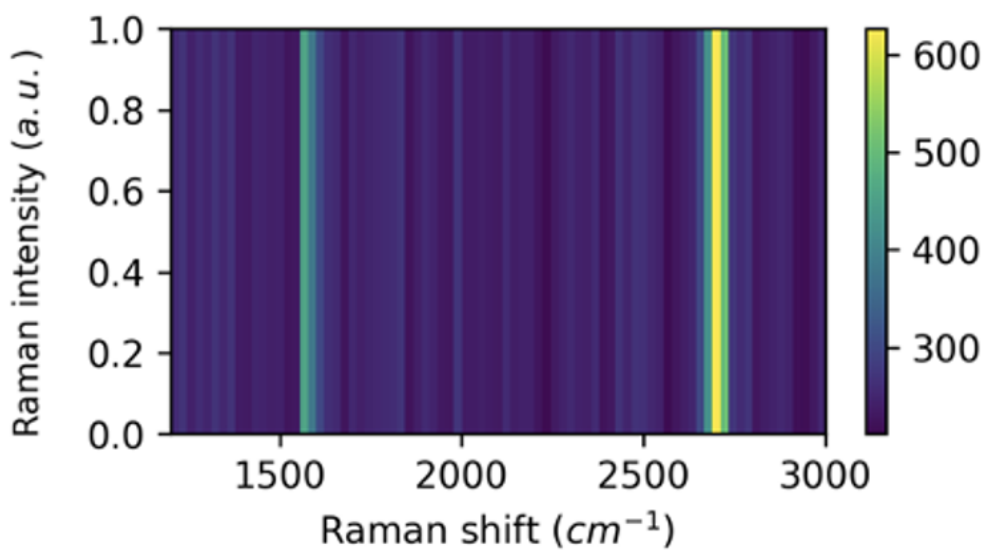

Figure 14 Visualization of activation mapping for the fourth convolutional layer of the classifier under an input spectrum Signal-to-Noise Ratio (SNR) of 0.

## 5. Summary

This study delves into the utilization of deep learning methodologies, encompassing classification-based and generative approaches, to enhance Raman spectroscopy analysis, with a specific emphasis on expediting the identification of 2D materials. The research addresses the classification of diverse twist angles inherent in bilayer graphene. Operating within the framework of machine learning for intricate data processing, deep learning techniques are employed to discern complex relationships inherent in experimentally obtained Raman spectroscopy data. The pivotal contribution of this research involves the development of a GAN-CNN model, comprising generator, discriminator, and classifier modules. This study expounds on the model's training outcomes and performance metrics. Comparative analyses with traditional machine learning algorithms and alternative deep learning networks validate the superiority of the proposed model in terms of classification accuracy. Additionally, the study simulates real experimental conditions by introducing noise to Raman spectra, evaluating the robustness and resistance to interference of the GAN-CNN model. The outcomes substantiate that the proposed deep learning model surpasses alternative approaches in its resilience to noise.

**CRediT authorship contribution statement**

**Dan Hu:** Investigation, Methodology, Software, Validation, Writing-Original draft, Writing-reviewing and Editing. **Ting-Fung Chung:** Data Curation, Writing-reviewing and Editing. **Yong P. Chen:** Conceptualization, Methodology, Project Administration, Resources, Funding acquisition, Supervision, Writing-reviewing and Editing. **Yaping Qi:** Conceptualization, Investigation, Methodology, Project Administration, Resources, Funding acquisition, Software, Validation, Supervision, Writing-Original draft, Writing-reviewing and Editing.

**Data availability**

The data are available from the corresponding authors upon reasonable request.

**Code availability**

The codes are available from the corresponding authors upon reasonable request.

**Acknowledgements**

We acknowledge support of this work by Macau Science and Technology Development Fund (FDCT Grants 0031/2021/ITP), Purdue University Discovery Park Big Idea Challenge program, JSPS KAKENHI (No. 22H00278), Tohoku University TUMUG Startup Research Fund, and AIMR Overseas dispatch program for young researchers FY2023.

## Conflict of interest

The authors declare that they have no conflict of interest.

## References

[1] V. Meunier, M. C. Ovín Ania, A. Bianco, Y. Chen, G. B. Choi, Y. A. Kim, N. Koratkar, C. Liu, J. M. Díez Tascón, and M. Terrones, "Carbon science perspective in 2022: Current research and future challenges," Carbon, 2022.

[2] Y. Qi, M.A. Sadi, D. Hu, M. Zheng, Z. Wu, Y. Jiang, Y.P. Chen, Recent Progress in Strain Engineering on Van der Waals 2D Materials: Tunable Electrical, Electrochemical, Magnetic, and Optical Properties, Advanced Materials 35(12) (2023) 2205714.

[3] I. Childres, Y. Qi, M.A. Sadi, J.F. Ribeiro, H. Cao, Y.P. Chen, Combined Raman Spectroscopy and Magneto-Transport Measurements in Disordered Graphene: Correlating Raman D Band and Weak Localization Features, Coatings 12(8) (2022) 1137.

[4] M. Yankowitz, S. Chen, H. Polshyn, Y. Zhang, K Watanabe, T Taniguchi, D. Graf, A. F. Young, and C. R. Dean, "Tuning superconductivity in twisted bilayer graphene," Science, vol. 363, no. 6431, pp. 1059–1064, 2019.

[5] A. L. Sharpe, E. J. Fox, A. W. Barnard, J. Finney, K. Watanabe, T. Taniguchi, M. Kastner, and D. Goldhaber-Gordon, "Emergent ferromagnetism near three-quarters filling in twisted bilayer graphene," Science, vol. 365, no. 6453, pp. 605–608, 2019.

[6] B. Padhi, C. Setty, and P. W. Phillips, "Doped twisted bilayer graphene near magic angles: Proximity to wigner crystallization, not mott insulation," Nano letters, vol. 18, no. 10, pp. 6175–6180, 2018.

[7] Y. Cao, D. Rodan-Legrain, O. Rubies-Bigorda, J. M. Park, K. Watanabe, T. Taniguchi, and P. Jarillo-Herrero, "Tunable correlated states and spin-polarized phases in twisted bilayer–bilayer graphene," Nature, vol. 583, no. 7815, pp. 215–220, 2020.

[8] C. N. Lau, M. W. Bockrath, K. F. Mak, and F. Zhang, "Reproducibility in the fabrication and physics of moiré materials," Nature, vol. 602, no. 7895, pp. 41–50, 2022

[9] N. Sheremetyeva, M. Lamparski, C. Daniels, B. Van Troeye, and V. Meunier, "Machine-learning models for raman spectra analysis of twisted bilayer graphene," Carbon, vol. 169, pp. 455–464, 2020.

[10] P. Solís-Fernández and H. Ago, "Machine learning determination of the twist angle of bilayer graphene by raman spectroscopy: Implications for van der waals heterostructures," ACS Applied Nano Materials, vol. 5, no. 1, pp. 1356–1366, 2022.

[11] T. Vincent, K. Kawahara, V. Antonov, H. Ago, and O. Kazakova, "Data cluster analysis and machine learning for classification of twisted bilayer graphene," Carbon, vol. 201, pp. 141–149, 2023.

[12] Y. Qi, D. Hu, Y. Jiang, Z. Wu, M. Zheng, E.X. Chen, Y. Liang, M.A. Sadi, K. Zhang, Y.P. Chen, Recent Progresses in Machine Learning Assisted Raman Spectroscopy, Advanced Optical Materials (2023) 2203104.

[13] Y. Qi, D. Hu, Z. Wu, M. Zheng, G. Cheng, Y. Jiang, Y.P. Chen, Deep Learning Assisted Raman Spectroscopy for Rapid Identification of 2D Materials, https://doi.org/10.48550/arXiv.2312.01389.

[14] C. Muehlethaler, M. Leona, and J. R. Lombardi, "Review of surface enhanced raman scattering applications in forensic science," Analytical Chemistry, vol. 88, no. 1, pp. 152–169, 2016.

[15] L. Alzubaidi, J. Zhang, A. J. Humaidi, A. Al-Dujaili, Y. Duan, O. Al-Shamma, J. Santamaría, M. A. Fadhel, M. Al-Amidie, and L. Farhan, "Review of deep learning: Concepts, cnn architectures, challenges, applications, future directions," Journal of big Data, vol. 8, pp. 1–74, 2021.

[16] J. Wang, L. Perez, et al., "The effectiveness of data augmentation in image classification using deep learning," Convolutional Neural Networks Vis. Recognit, vol. 11, no. 2017, pp. 1–8, 2017.

[17] I. Goodfellow, J. Pouget-Abadie, M. Mirza, B. Xu, D. Warde-Farley, S. Ozair, A. Courville, and Y. Bengio, "Generative adversarial networks," Communications of the ACM, vol. 63, no. 11, pp. 139–144, 2020.

[18] K. Wang, C. Gou, Y. Duan, Y. Lin, X. Zheng, and F.-Y. Wang, "Generative adversarial networks: Introduction and outlook," IEEE/CAA Journal of Automatica Sinica, vol. 4, no. 4, pp. 588–598, 2017.

[19] M. Arjovsky, S. Chintala, and L. Bottou, "Wasserstein generative adversarial networks," in international conference on machine learning, PMLR, 2017, pp. 214–223.

[20] L. Van der Maaten and G. Hinton, "Visualizing data using t-sne.," Journal of machine learning research, vol. 9, no. 11, 2008.

[21] R. He, T.-F. Chung, C. Delaney, C. Keiser, L. A. Jauregui, P. M. Shand, C. Chancey, Y. Wang, J. Bao, and Y. P. Chen, "Observation of low energy raman modes in twisted bilayer graphene," Nano letters, vol. 13, no. 8, pp. 3594–3601, 2013.